\documentclass{ifacconf}

\usepackage{graphicx}      % include this line if your document contains figures
\usepackage{natbib}        % required for bibliography
\usepackage{bm}
\usepackage{amssymb}
\usepackage{amsmath}
\usepackage{color}
\usepackage{graphicx}
\usepackage{mathtools}
\usepackage{url}
\usepackage{csquotes}
\DeclareMathOperator*{\argmin}{argmin}
\usepackage{algorithm}
\usepackage[noEnd=true]{algpseudocodex} % Makes vertical lines instead of "ends"

\newtheorem{lemma}{Lemma}
\usepackage[table]{xcolor}

\usepackage{mathtools}
\DeclarePairedDelimiter{\ceil}{\lceil}{\rceil}
\begin{document}
\begin{frontmatter}

\title{Gauss-Newton accelerated MPPI Control}%\thanksref{footnoteinfo}} 

%\thanks[footnoteinfo]{Sponsor and financial support acknowledgment
%goes here. Paper titles should be written in uppercase and lowercase
%letters, not all uppercase.}

\author[First]{Hannes Homburger} 
\author[Second]{Katrin Baumgärtner} 
%\author[First]{Patrick Hoher}
\author[Third]{Moritz Diehl}
\author[First]{Johannes Reuter}

\address[First]{Institute of System Dynamics, HTWG Konstanz - University of Applied Sciences, 78462 Konstanz, Germany
        hhomburg@htwg-konstanz.de}
\address[Second]{Department of Microsystems Engineering (IMTEK),
    	University of Freiburg, 79110 Freiburg, Germany}
\address[Third]{Department of Microsystems Engineering (IMTEK) and Department of Mathematics, University of Freiburg, \\79110 Freiburg, Germany}

\begin{abstract}                % Abstract of not more than 250 words.
Model Predictive Path Integral (MPPI) control is a sampling-based optimization method that has recently attracted attention, particularly in the robotics and reinforcement learning communities. MPPI has been widely applied as a GPU-accelerated random search method to deterministic direct single-shooting optimal control problems arising in model predictive control (MPC) formulations. MPPI offers several key advantages, including flexibility, robustness, ease of implementation, and inherent parallelizability. However, its performance can deteriorate in high-dimensional settings since the optimal control problem is solved via Monte Carlo sampling.
To address this limitation, this paper proposes an enhanced MPPI method that incorporates a Jacobian reconstruction technique and the second-order Generalized Gauss-Newton method. This novel approach is called \textit{Gauss–Newton accelerated MPPI}. The numerical results show that the Gauss-Newton accelerated MPPI approach substantially improves MPPI scalability and computational efficiency while preserving the key benefits of the classical MPPI framework, making it a promising approach even for high-dimensional problems. 
\end{abstract}


\end{frontmatter}
%===============================================================================

\section{Introduction}

Model Predictive Path Integral (MPPI) control is a sampling-based optimal feedback control method widely used in robotics applications \citep{Kazim2024}. Its theoretical foundation and its name originate from input-affine stochastic optimal control formulations, where the solution can be expressed as a \textit{path integral} derived from the Feynman–Kac lemma \citep{Kappen2005}. Via an \textit{information-theoretic} approach, the framework has been generalized to nonlinear dynamical systems \citep{Williams2018} and is commonly used as a GPU-accelerated random search variant for deterministic optimal control problems, particularly arising in nonlinear model predictive control (NMPC), see \cite{Zhang.2024} and \cite{Halder2025}. % Yan.2024,

To model and simulate complex physical systems, simulation engines such as  %\texttt{Brax} \citep{freeman_brax_2021},
\texttt{Dart} \citep{lee_dart_2018}, \texttt{MuJoCo} \citep{todorov_mujoco_2012}, and \texttt{Isaac Sim} by \texttt{NVIDIA} are widespread tools.
However, these engines often do not provide sensitivity information.
Given the black-box nature of specific simulation engines, the MPPI control method offers desirable properties, including flexibility and robustness. %ease of implementation, and inherent parallelizability. 
These properties are exploited, e.g., by \cite{Sundaralingam.2023} to solve optimal control problems using a simulation engine to represent the dynamics. Nevertheless, the performance of MPPI deteriorates in high-dimensional settings because the underlying sampling distribution can exhibit high variance, which can lead to convergence issues, particularly when solving optimal control problems with long time horizons or unstable system dynamics.

To mitigate this limitation, various approaches have been proposed to enhance sampling efficiency. For example, learning more informative sampling distributions can increase the proportion of low-cost samples \citep{sacks2023learning}. Furthermore, the use of learned basis skills within the MPPI framework has been investigated in \cite{Homburger2022.b} and later extended in \cite{Trevisan2024}, which incorporated general ancillary controllers. Classical optimization methods have been applied to enhance sampling-based controllers. For example, path integral control has been enhanced through differential dynamic programming approaches ~\citep{Lefebvre.2019} and an uncertainty-aware MPPI variant has been leveraged with iterative linear–quadratic Gaussian control~\citep{Gandhi.2021}.  A review of various developments is provided in \cite{Kazim2024}. Although such methods mitigate sampling inefficiency to some extent, they remain fragile to the curse of dimensionality, manifested via the variance of the Monte Carlo estimator.

In the field of numerical optimization, second-order techniques such as the Generalized Gauss-Newton method \citep{Schraudolph.2002} are commonly employed to approximate the Hessian matrix within Newton-type optimization schemes that can be highly efficient and exploit the problem's structure \citep[Sec. 8.6]{Rawlings2022}. In particular, the favorable local convergence properties of the Generalized Gauss-Newton method have been investigated by \cite[Sec. 2.3]{Messerer.2021}.

\textbf{Contribution.} In this work, we integrate Jacobian reconstruction techniques and the second-order Generalized Gauss-Newton method from numerical optimization into the MPPI control framework to improve its performance and scalability. The presented approach preserves the key properties of classical MPPI, given by its flexible interface and the parallel evaluation of the objective.  

\textbf{Structure.} The remainder of this paper is organized as follows. Section \ref{sec:background} formulates the optimal control problem to be solved. Section \ref{sec:mppi} provides a review of the deterministic MPPI control approach. The proposed Gauss-Newton-accelerated MPPI method is introduced in Section \ref{sec:GN_MPPI} and compared with other optimization approaches in Section~\ref{sec:exp}. Finally, Section \ref{sec:con} concludes the paper.

\section{Problem Formulation}\label{sec:background}
Let us consider a deterministic black-box simulator $F:\mathbb{R}^{n_x}\times\mathbb{R}^{Nn_u}\rightarrow\mathbb{R}^{(N+1)n_x} $, which might be highly nonlinear or even nonsmooth, an initial state $\overline x_0\in\mathbb{R}^{n_x}$, and a discrete-time $N$-step input trajectory $U=[u_0^\top,u_1^\top,\ldots,u_{N-1}^\top]^\top\in\mathbb{R}^{N n_u}$, then the state trajectory can be computed by
\begin{equation}\label{eq_simulator}
    X=F(\overline x_0,U),
\end{equation}
where $X=[x_0^\top,x_1^\top,\ldots,x_{N}^\top]^\top\in\mathbb{R}^{(N+1) n_x}$ contains the simulated states $x_k\in\mathbb{R}^{n_x}$ for $k=0,1,...,N$.

%deterministic discrete-time system dynamics $f:\mathbb{R}^{n_x}\times \mathbb{R}^{n_u} \rightarrow \mathbb{R}^{n_x}$, an initial state $\overline x_0\in\mathbb{R}^{n_x}$, an $N$-step input trajectory $U=[u_0^\top,u_1^\top,\ldots,u_{N-1}^\top]^\top\in\mathbb{R}^{N n_u}$ with elements $u_k\in\mathbb{R}^{n_u}$ for $k=0,1,...,N-1$, then the state trajectory can be determined by forward simulation 
%\begin{subequations}
%\begin{align}
%    x_0(U;\overline x_0)&= \overline x_0,\label{eq_x_init}\\
%    x_{k+1}(U;\overline x_0)&=f(x_k(U;\overline x_0),u_k,w_k)\label{eq_x}
%\end{align}    
%\end{subequations}
%with states $x_k\in\mathbb{R}^{n_x}$ for $k=0,1,...,N$.
Given an initial state $\overline x_0$, the performance of a control trajectory $U$ is described by the \textit{overall costs} given by
\begin{equation}\label{eq_objective}
       C(\overline x_0,U):=\sum_{k=0}^{N-1} L_k(x_k(\overline x_0,U),u_k)+E(x_N(\overline x_0,U)),  
\end{equation}
where $L_k:\mathbb{R}^{n_x}\times\mathbb{R}^{n_u}\rightarrow \mathbb{R}^+_0$ for $k=0,\ldots,N-1$ are the stage costs and $E:\mathbb{R}^{n_x}\rightarrow \mathbb{R}^+_0$ is the terminal cost.

The optimal input trajectory for the corresponding deterministic, unconstrained, finite-horizon, and discrete-time optimal control problem
is given by
\begin{equation}\label{eq_NLP}
U^\star(\overline x_0)=\argmin_{U\in\mathbb{R}^{N {n_u}}}C(\overline x_0,U),    
\end{equation}
where $U^\star:\mathbb{R}^{n_x}\rightarrow\mathbb{R}^{N n_u}$ is the optimal input trajectory. 

 We can express the objective \eqref{eq_objective} in the special structure  
\begin{equation}\label{eq:MPPI_convex_over_non}
     C(\overline x_0,U)= \Phi(R(\overline x_0,U)), 
  \end{equation}
  such that the objective is a composition of a nonlinear, unknown, and possibly nonsmooth \textit{inner function} $R:\mathbb{R}^{n_x}\times\mathbb{R}^{N n_u}\rightarrow \mathbb{R}^{n_R}$ and a  \textit{outer function} $\Phi:\mathbb{R}^{n_R}\rightarrow \mathbb{R}^{+}_0$. In case of $L_k$ for all $k=0,\ldots, N-1$ and $E$ convex, the outer function $\Phi$ is convex, and we obtain a \textit{convex-over-unknown} cost structure in Eq.~\eqref{eq:MPPI_convex_over_non}.

\begin{figure}[t]
  \centering
  \medskip
  \includegraphics[width=\linewidth]{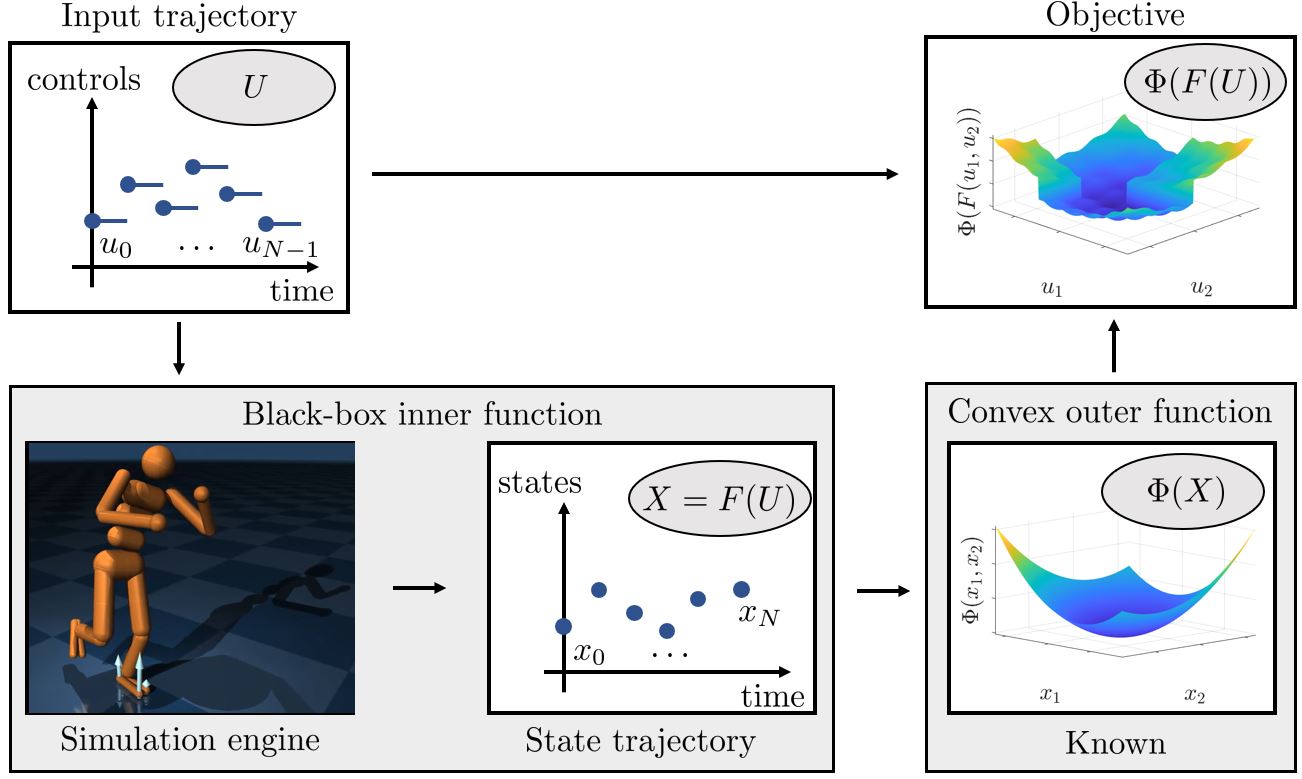}
  \caption{Schematic visualization of the \textit{convex-over-unknown} structure of the NLP objective in Eq.~\eqref{eq:MPPI_convex_over_non}. For simplicity, the special case $R=X$ is considered.}
  \label{fig:convex_over_nonlinear}
\end{figure}

Note that the convex-over-unknown structure is typical in NMPC approaches, where the system dynamics are simulated using a black-box engine and the cost function is chosen to be convex.
This structure is depicted in Figure~\ref{fig:convex_over_nonlinear}.

\section{Deterministic MPPI Control}\label{sec:mppi}
To solve the optimal control problem \eqref{eq_NLP} numerically, the deterministic MPPI approach is based on the iterates $U_{k+1}=U_k+\Delta U_\mathrm{MPPI}(U_k)$ with step $\Delta U_\mathrm{MPPI}(U_k)= $
\begin{equation}\label{eq_MPPI}
    \frac{1}{\eta_k}\mathbb{E}_{W\sim\mathcal{N}(0, \Sigma_k)}\left[W\exp\left( -\frac{1}{\lambda_k} C(\overline x_0,U+W) \right)\right],
\end{equation}
where $U_k\in\mathbb{R}^{N n_u}$ is an initial guess, $W\in\mathbb{R}^{N n_u}$ is a multivariate normal distributed random variable with mean zero and covariance $\Sigma_k$, the scalar \begin{equation}
    \eta_k=\mathbb{E}_{W\sim\mathcal{N}(0, \Sigma_k)}\left[\exp\left( -\frac{1}{\lambda_k} C(\overline x_0,U+W) \right)\right]
\end{equation}
  normalizes the distribution, and $\lambda_k>0$ is a scalar tuning parameter, see e.g. the presentation by \cite{Yi.2024}. 
For the limit $\lambda_k\rightarrow 0$, the convergence of the deterministic MPPI method can be proven using Laplace's method; see e.g. Thm.~1 in \cite{Homburger.2025} for a similar result. The basis for the application of the MPPI method is to approximate the expectation in Eq.~\eqref{eq_MPPI} by the empirical mean calculated by a finite number of $M$ samples
\begin{equation}
    \label{eq_MPPI_approx}
    \Delta U_{\mathrm{MPPI}}(U_k) \approx  \frac{\sum_{m=1}^M\left[W_m\;\omega_m \right]}{\sum_{m=1}^M\left[\omega_m\right]},
\end{equation}
where 
\begin{equation}\label{eq_weight}
    \omega_m:=\exp\left( -\frac{1}{\lambda_k} C(\overline x_0,U_k+W_m) \right)
\end{equation} 
are the \textit{sample weights} for $m=1,\ldots, M$ samples. Note that the presented deterministic MPPI approach in Eq.~\eqref{eq_MPPI} is only one of many methods within the MPPI framework \citep{Kazim2024}. The essence of this framework is that the empirical weighted mean in Eq.~\eqref{eq_MPPI_approx} underlies most MPPI algorithms (cf., e.g., \cite{Williams2018,Kazim2024,Halder2025}). 
The samples and corresponding weights in Eq.~\eqref{eq_MPPI_approx} can be generated in parallel with algorithms tailored to GPUs \citep{Vlahov.2024}. In general, MPPI does not require any sensitivity information. This allows MPPI to treat problems in a black-box setting. %However, the question arises of how many samples are necessary for a sufficient approximation.
\subsection{Sample complexity analysis}
The motivation for the approximation of the expectation by a weighted empirical mean in Eq.~\eqref{eq_MPPI_approx} is based on the \textit{strong law of large numbers} \citep[Thm.~1.3.1]{Vershynin.2018}.
Now the question arises: how many samples are necessary to provide a good estimate using Eq.~\eqref{eq_MPPI_approx}? The following lemma is a standard result from Monte Carlo theory and provides a bound on the expected error in general Monte Carlo approaches that is dependent on the variance of the random variable and the number of samples in the empirical mean. Lemma~\ref{theory:lemma_MC_error_bound} is adapted from Chapter~3 of the textbook by \cite{Vershynin.2018}.
\begin{lemma}[Monte Carlo error bound]\label{theory:lemma_MC_error_bound}
    Let $X_1,X_2,\ldots,X_n$ be i.i.d. random vectors in $\mathbb{R}^{n_x}$ with mean $\mathbb{E}[X]=\mu$ and finite covariance matrix $\Sigma\succ0$. Then for the empirical mean of $n$ samples $\overline X_n:=\frac{1}{n}\sum_{i=1}^n X_i,$ it follows
    \begin{equation*}
        \mathbb{E}\biggl[\,
\left\lVert \overline X_n - \mu \right\rVert_2
\biggr]\leq \frac{\sqrt{ \mathrm{tr}(\Sigma)}}{\sqrt{n}}.
    \end{equation*}
\end{lemma}
%\begin{figure}[b]
%  \centering
%  \medskip
%  \includegraphics[width=1\linewidth]{figures/mean_error_paper.pdf}
%  \vspace{-0.3cm}
%  \caption{Mean error over sample number for different dimensions of the random variable. The empirical mean error decreases with a slope $\propto 1/\sqrt{n}$ independent of the dimension $n_x$ of the random variable.}
%  \label{fig:theory_mean_error}
%\end{figure}

\textit{\textbf{Sketch of proof:}}
Let $Y=\overline X_n-\mu$ and note that $\mathrm{Cov}(Y)=\frac{\Sigma}{n}$. The Jensen inequality for the Euclidean norm yields \[ \mathbb{E}\left[\,\left|\left| Y\right|\right|_2\,\right]\leq  \sqrt{\mathbb{E}\left[\left|\left| Y\right|\right|_2^2\right]}. \]
By identifying 
\[\mathbb{E}\left[\left|\left| Y\right|\right|_2^2\right]=\mathrm{tr(Cov(Y))}=\mathrm{tr}\left(\frac{\Sigma}{n}\right)=\frac{\mathrm{tr}(\Sigma)}{n},\]
  the claimed result is immediate.\qed

Note that the trace can be bounded by $\mathrm{tr}(\Sigma)\leq n_x\overline \lambda$, where $\overline{\lambda}$ is the largest eigenvalue of $\Sigma$, which coincides with the largest diagonal entry if $\Sigma$  is diagonal.
 The \textit{strong law of large numbers} motivates the use of the empirical mean, and Lemma~\ref{theory:lemma_MC_error_bound} quantifies the number of samples required to achieve a reliable approximation on average. Because the error bound scales as $\propto 1/\sqrt{n}$, obtaining an additional correct decimal digit requires increasing the sample size by a factor of 100. This convergence rate is slow, particularly for low-dimensional problems. The Monte-Carlo method's \enquote{reason to be} lies in its straightforward applicability to multivariate random variables, %as illustrated in Figure~\ref{fig:theory_mean_error}, 
 where the decrease of the error bound holds regardless of dimensionality.
 
 Note that tighter bounds can be established for specific applications of MPPI control \citep{Yoon.2022}, e.g., in the linear quadratic regulator (LQR) case via Hoeffding’s inequality, the sample size required exhibits a logarithmic dependence
on the dimension of the control input trajectory \citep[Sec. 4.A]{Patil.2024}.  

\subsection{Properties of the MPPI algorithm}
The MPPI solution Eq.~\eqref{eq_MPPI_approx} relies solely on evaluations of the objective 
$C$, which embeds the simulator’s black-box dynamics, requiring no gradient or Hessian information and making it easy to implement. This allows MPPI to handle problems where sensitivities are unavailable.  MPPI is also robust to poor initial guesses, as even a single low-cost sample can guide the update. See \cite{Kazim2024} for a detailed discussion. 

Despite the presented benefits of MPPI approaches, the algorithm completely ignores the internal problem structure of the optimal control problem and first- and second-order information that is valuable to exploit in most problem formulations \citep[Sec. 8]{Rawlings2022}.
Exploiting the structure of the problem and the information of first and second-order is a key element to achieve fast convergence in Newton-type optimization algorithms tailored to the problem structure \citep{Baumgartner.2023}. 

Facing the different approaches, the question arises whether it is possible to combine the benefits of both methods and achieve faster convergence by using structured evaluations of the cost function rather than purely sampling.
%In the following section, we present Gauss-Newton accelerated MPPI.
%\begin{remark}
%The name \textit{model predictive path integral} control is widely used and well-known in the robotics and reinforcement learning community, but can be misleading in the given setting.
%\end{remark}

\section{Gauss-Newton accelerated MPPI}\label{sec:GN_MPPI}
In this section, we introduce the Gauss-Newton accelerated MPPI method.
We start by reviewing techniques to reconstruct Jacobian information from black-box functions.
Then, the Gauss-Newton accelerated MPPI algorithm is presented.
In the remainder of this paper, we omit the parameter dependency on the initial state in the notation, e.g., we will use $C(U)\equiv C(\overline x_0,U)$ and $R(U)\equiv R(\overline x_0,U)$ for readability.
\subsection{Jacobian reconstruction}
The Generalized Gauss-Newton approach is based on the Jacobian of the unknown residuum function $R$. To reconstruct Jacobian information from black-box functions, several approaches have been proposed in the literature. We select the \textit{Gaussian smoothing} method because it relies on a similar Gaussian-distributed function evaluation as standard MPPI, and it is applicable even if the unknown inner function is possibly nonsmooth.
%\subsubsection{Jacobian reconstruction by finite differences}
% To reconstruct the sensitivity information, we exploit the fact that for any twice differentiable function
%\begin{equation}\label{eq:fd}
%    \frac{R(U_k+tp)-R(U_K)}{t}=\frac{\partial R}{\partial U} (U_k) p+\mathcal{O}(t)
%\end{equation}
%hold (cf. \cite[Sec. 2.1]{Berahas.2022}). That means that we can get the directional derivative by a small perturbation $t>0$ in the direction $p$ with $\lvert \lvert p \lvert \lvert_1=1$. The overall computational complexity for the Jacobian reconstruction is 
%\begin{equation}
%\mathrm{cost}\left(\frac{\partial R}{\partial U}(U_k) \right)=(n_{N\:n_u}+1)\cdot \mathrm{cost}(R(U_k)),     
%\end{equation}
%where $\mathrm{cost}(R)$ denotes the computational cost to evaluate the function $R$. Note that the function evaluations can be done in parallel, e.g., on a GPU, such that under the assumption of ideal parallelization 
%\begin{equation}
%\mathrm{cost\_GPU}\left(\frac{\partial R}{\partial U}(U_k) \right)= %\mathrm{cost}(R(U_k))    
%\end{equation}
%follows.
\subsubsection{Jacobian reconstruction by Gaussian smoothing.}
\begin{figure}[t]
  \centering
  \medskip
  \includegraphics[width=\linewidth]{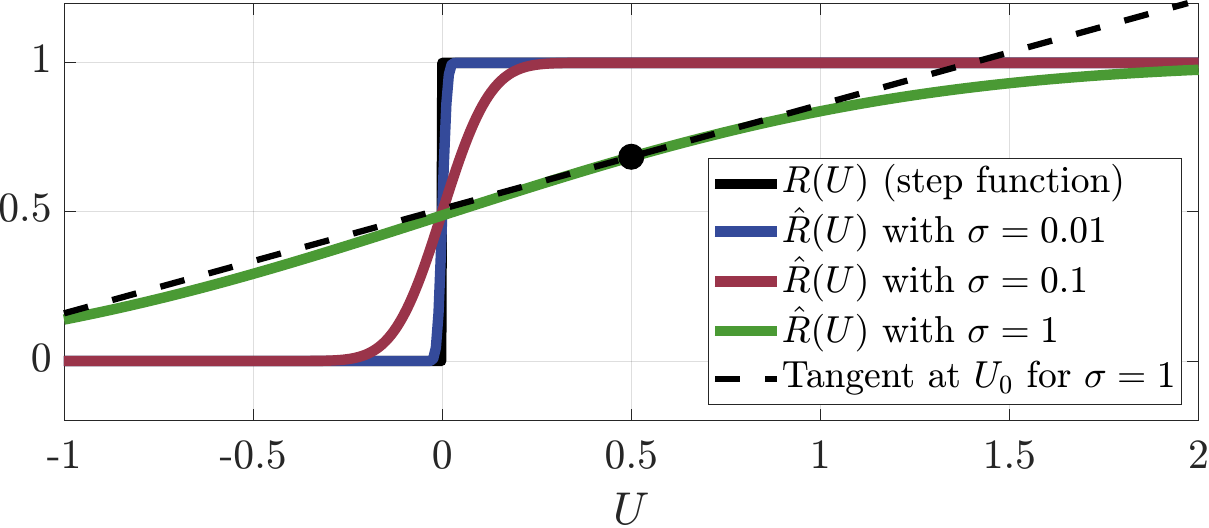}
  \vspace{-0.3cm}
  \caption{Heaviside step function and Gaussian smoothed approximations, here scalar $\Sigma=\sigma^2$ with $\sigma>0$.}
  \label{fig:gaussian_smoothing}
\end{figure}

To approximate the Jacobian of the unknown inner function $R$, the function $R$ can be treated by \textit{Gaussian smoothing}, resulting in a smooth function
\begin{equation}
    \hat R(U_k)=\mathbb{E}_{W\sim\mathcal{N}(0,\Sigma)}[R(U_k+W)],
\end{equation}
 where $\Sigma\succ0$ is a diagonal positive definite covariance matrix (cf. Sec.~2.3 in \cite{Berahas.2022}). The influence of the covariance is depicted in Figure~\ref{fig:gaussian_smoothing}. By exploiting properties of derivatives of the expectation \citep{Asmussen.2007}, we obtain the Jacobian as expectation
%\begin{equation}\label{eq_Gaussian_smoothing_jacobian}
    %\frac{\partial\hat R}{\partial U}(U_k)=\frac{1}{\sigma}\mathbb{E}_{W\sim\mathcal{N}(0, I)} \left[R(U_k+\sigma W) W^\top \right].
%\end{equation}
\begin{equation}\label{eq_Gaussian_smoothing_jacobian}
    \frac{\partial\hat R}{\partial U}(U_k)=\mathbb{E}_{W\sim\mathcal{N}(0, \Sigma)} \left[R(U_k+ W) W^\top \right]\:\Sigma^{-1}.
\end{equation}
By assuming the existence of the Jacobian $\tfrac{\partial R}{\partial U}(U_k)$, it can be  recovered by $\tfrac{\partial \hat R}{\partial U}(U_k)$ for $||\Sigma|| \rightarrow 0$  (cf. Eq.~(2.10) in \cite{Berahas.2022}). 
The expectation can be approximated by Monte Carlo sampling with $M\in\mathbb{N}$ samples.
This results in the computational complexity  
 \begin{equation}
\mathrm{cost}\left(\frac{\partial\hat R}{\partial U}(U_k) \right)\approx M\cdot \mathrm{cost}( R(U_k))     
\end{equation}
that can be accelerated by execution in parallel, such that
\begin{equation}\label{eq_GPU_cost}
\mathrm{cost}_\mathrm{GPU}\left(\frac{\partial \hat R}{\partial U}(U_k) \right)\approx \ceil[\bigg]{\frac{M}{M_\mathrm{GPU}}}  \cdot \mathrm{cost}(R(U_k)) 
\end{equation}
follows under the hard assumption of ideal parallelization of $M_\mathrm{GPU}\in\mathbb{N}$ processing units. We can approximate the Jacobian $J(U_k):=\frac{\partial \hat R}{\partial U}(U_k)$ of the smoothed \textit{inner} black-box function $R$ of \eqref{eq:MPPI_convex_over_non}.  Note that in the case of a differentiable inner function $R$, finite differences \citep{Curtis.1974} is another approach to approximate the Jacobian.
%We now use the Jacobian within a Generalized Gauss-Newton approach.

\subsection{Generalized Gauss-Newton (GGN) method}
The Generalized Gauss-Newton (GGN) method \citep{Schraudolph.2002} is an iterative second-order optimization approach applicable to costs with a convex-over-nonlinear structure, as in Eq.~\eqref{eq:MPPI_convex_over_non}, which is commonly used in NMPC applications.

Within the GGN method, the next iterate is computed as 
\begin{equation}\label{eq:MPPI_GGN_Step}
    U_{k+1}=U_k\underbrace{-B_\mathrm{GGN}(U_k)^{-1}\;\nabla[ \Phi(R(U_k))]}_{=:\Delta_{\mathrm{GGN}}(U_k)},
\end{equation}
that is the minimizer of the convex quadratic \textit{GGN subproblem}
\begin{equation*}
    \argmin_{U_\in\mathbb{R}^{n_u}}  \nabla[ \Phi(R)]^\top(U-U_k)+\tfrac{1}{2}(U-U_k)^\top B_\mathrm{GGN}(U-U_k),
\end{equation*}
where $\nabla [\Phi(R)]^\top=J^\top\nabla[\Phi](R)$ and the GGN Hessian is 
\begin{equation*}
    B_\mathrm{GGN}(U_k):=J(U_k)^\top\: \nabla^2  \Phi(R(U_k))\: J(U_k)\succ0.
\end{equation*}
The error of the GGN Hessian approximation with respect to the exact Hessian is
\[\nabla^2[\Phi(R(U))]-B_\mathrm{GGN}(U)=\sum_{j=1}^{n_R}\nabla^2 R_j(U)\nabla_{R_j}\Phi(R(U)),\]
and contains the potentially indefinite terms of the Hessian of the original objective.
Fast convergence is achieved when the curvature of the inner functions is small. See \cite[Sec. 2.3]{Messerer.2021} for a detailed investigation of the convergence properties of the GGN method. 

\begin{algorithm}[b]
    \caption{Gauss-Newton accelerated MPPI}
    \label{alg:GNaccMPPI}
    {\small
    \begin{algorithmic}[1]
        \setlength{\itemsep}{1pt}
        \Require Initial guess $U_0$, maximum number of iterations $K$, number of samples $M$, 
                 initial covariance $\Sigma_0$, shrinking rate $\beta$, step-length set $\mathcal{A}$
        \For{$k \in \{0,1,\dots,K-1\}$}
          %  \For{$m \in \{1,\dots,M\}$} \Comment{In parallel}
           %     \State $W_m \sim \mathcal{N}(0,I)$ \Comment{Generate samples}
            %    \State $\omega_m \gets R(U_k + \sigma_k W_m)$ 
             %       \Comment{Compute weight \eqref{eq_Gaussian_smoothing_jacobian}}
            %\EndFor
            \State $J_k \gets \textsc{GetJacobian}(U_k,\Sigma_k,M)$ 
                \Comment{Parallel smoothing \eqref{eq_Gaussian_smoothing_jacobian}}
            \State $\Delta_{\mathrm{GGN}} \gets 
                -B_{\mathrm{GGN}}(U_k)^{-1}\,\nabla \Phi(R(U_k))$ 
                \Comment{Full GGN step \eqref{eq:MPPI_GGN_Step}}
            \State $\alpha^\star \gets \textsc{LineSearch}(U_k,\Delta_{\mathrm{GGN}},\mathcal{A})$
                \Comment{Parallel sol. to \eqref{eq:MPPI_length}}
                \State  $U^{k+1} \gets \textsc{GGNstep}(U_k,\Delta_{\mathrm{GGN}},\alpha^\star)$ \Comment{Apply GGN step \eqref{eq:MPPI_ad_GGN_Step}}
                \State $\Sigma_{k+1} \gets \beta\:\Sigma_k$ \Comment{Shrink covariance}
        \EndFor
        \State \Return $U^\star \gets U_K$ \Comment{Solution to \eqref{eq_NLP}}
    \end{algorithmic}
    }
\end{algorithm}

\subsection{Gauss-Newton accelerated MPPI algorithm}
Now the presented parts are put together to obtain the 
 Gauss-Newton accelerated MPPI method that is summarized in Algorithm~\ref{alg:GNaccMPPI}. 
 
 First, the Jacobian $J(U_k)$ is approximated using Gaussian smoothing, with $M$ parallel evaluations of the inner function $R$. Based on this approximation, the full GGN step is computed, after which, a standard line search method is employed to adjust the length of the GGN step \eqref{eq:MPPI_GGN_Step} for globalization.
This results in the iterative expression
\begin{equation}\label{eq:MPPI_ad_GGN_Step}
    U_{k+1}=U_k-\alpha^\star\; B_\mathrm{GGN}(U_k)^{-1}\;\nabla[ \Phi(R(U_k))],
\end{equation}
where the optimal step length $\alpha^\star$ is selected by
\begin{equation}\label{eq:MPPI_length}
    \alpha^\star=\argmin_{\alpha\in\mathcal{A}}C(U_k-\alpha\; B_\mathrm{GGN}(U_k)^{-1}\;\nabla[ \Phi(R(U_k))]),
\end{equation}
where $\mathcal{A}=\{\gamma^{k}|k=0,\ldots,M-1\}$ is a set containing $M<M_\mathrm{GPU}$ applicable step lengths with $\gamma\in(0,1)$. 
The computation of the optimal step length can be executed in parallel, leading to a computational cost
\begin{equation}
    \mathrm{cost}_\mathrm{GPU}\left(\alpha^\star\right)      \approx\mathrm{cost}(C(U_k))\approx\mathrm{cost}(R(U_k)).
\end{equation}
 Finally, the covariance used for Gaussian smoothing can be reduced to improve the accuracy of the Jacobian approximation in the subsequent iteration. 
 
 Computing the Jacobian in Eq.~\eqref{eq_GPU_cost} is necessary to determine the step direction in Eq.~\eqref{eq:MPPI_GGN_Step}, and the step length in Eq.~\eqref{eq:MPPI_length} can be optimized only after this direction has been obtained. Therefore, the overall computational cost of a Gauss-Newton accelerated MPPI step in parallel execution is roughly twice the cost of one evaluation of the inner function given in Eq.~\eqref{eq_simulator}.
The overall computational cost for $K$ steps in Algorithm~1 is roughly
 \begin{equation}\label{eq_GPU_Algo}
\mathrm{cost}_\mathrm{GPU}\left(\mathrm{Algo.~1} \right)\approx 2K\cdot \ceil[\bigg]{\frac{M}{M_\mathrm{GPU}}} \cdot \mathrm{cost}(R(U_k)).    
\end{equation}
Note that the computational cost estimates are derived under the assumption that the computational cost of the simulator Eq.~\eqref{eq_simulator} dominates that of all other algorithmic components.

\section{Numerical Experiments}\label{sec:exp}
To investigate the performance of the presented approaches, the following optimization problems are implemented in \textit{convex-over-unknown} fashion, i.e., the inner functions do not provide any sensitivity information. 

\begin{figure}[b]
  \centering
  \medskip
  \includegraphics[width=0.95\linewidth]{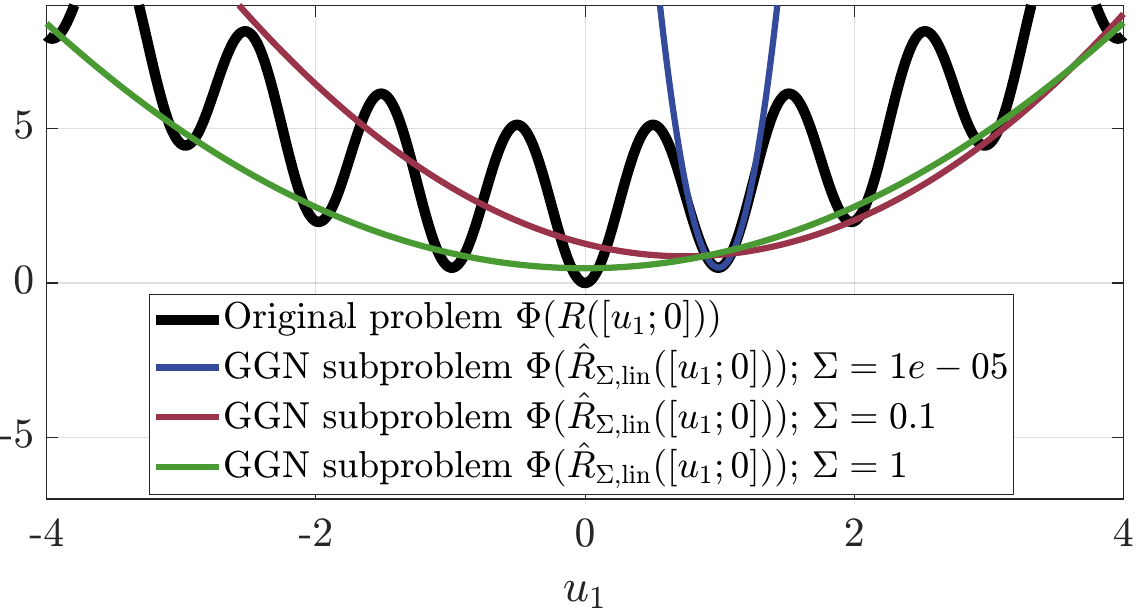}
  \vspace{-0.1cm}
  \caption{Visualization of Problem II with $u_2$ fixed at $0$. The GGN subproblems are expanded at $\hat u_1 = 0.9$. For small values of $\Sigma$, the GGN subproblem matches the original objective, while for larger $\Sigma$, the smoothed subproblem guides toward the global optimum.  }
  \label{fig:subproblem}
\end{figure}

\textbf{Problem I}: Rosenbrock function.
    The objective is 
    \[\Phi(R(U))=(1 - u_1)^2 + 100 \left(u_2 - u_1^2\right)^2\] with inner function $R(U)=[\sqrt 2(1 - u_1),\; \sqrt{200} \left(u_2 - u_1^2\right)]^\top$ and outer function $\Phi(R)=\tfrac12 RR^\top$. The decision variable is $U=(u_1,u_2)$ and we use the initial guess $ U_0=(0,0)$.\\
     \textbf{Problem II}: Rastrigin function.
    The objective is given by
    \[\Phi(R(U))=10+u_1^2-5\cos(2\pi u_1)+u_2^2-5\cos(2\pi u_2)\] with %inner function 
    %$R(U)=\left[u_1,\; u_2,\;(5-5\cos(2\pi u_1))^{\tfrac12},\;(5-5\cos(2\pi u_2))^{\tfrac12} \right]^\top$
    decision variable $U=(u_1,u_2)$. Problem II and the corresponding GGN subproblems are depicted in Figure~\ref{fig:subproblem}. We use the initial guess $U_0=(1.9,1.7).$\\
    \textbf{Problem III}: Heaviside function.
    The inner function is\[R(U)=\left\{\begin{array}{c}
           1, \quad \mathrm{for}\; u\geq0\\
           0, \quad \mathrm{for}\; u<0
    \end{array}\right.\] and the outer function is $\Phi=\tfrac{1}{2}R^2$ with decision variable $U=u\in\mathbb{R}$. We use the initial guess $U_0=0.5$.\\
    \textbf{Problem IV}: Linear optimal control of a double integrator.
    Discrete time dynamics $x_{k+1}=Ax_k+Bu_k$ for $k=0,...,N-1$ with $N\in\mathbb{N}$ and initial condition $x_0=\overline x_0\in\mathbb{R}^2$. The cost function is given by the quadratic costs
    $L(x_k,u_k)=x_k^\top Q x_k+ u_k^\top R u_k$
    and
    $E(x_N)=x_N^\top Q_N x_N$ with $Q,R,Q_N\succ 0$. We choose $N=50$ steps and $ U_0=0_{1\times 50}$.  \\ 
    \textbf{Problem V}: Nonlinear optimal control of the (nonsmooth) Furuta pendulum.
    The dynamics are given by a black-box implementation of Eq. (2) in \cite{Homburger.2025c}. The task is to track a reference trajectory over $N=20$ time steps, penalizing both tracking error and control effort. 
    We distinguish between \textit{Setting V.i}, where the standard dynamics are employed, and \textit{Setting V.ii}, where we add the simple nonsmooth friction model  \[
    u_{k,\mathrm{apply}}=\left\{\begin{array}{cl}
         0,\quad &\mathrm{for }\; |u_k|\leq {u}_\mathrm{friction}\;  \\
         u_k, \quad &\mathrm{else }
    \end{array} \right.\]
    at the input of the dynamical system. For both settings, the explicit RK4 method is used to discretize the smooth dynamics, and the nonsmooth friction model is applied to the discrete-time dynamics.

For comparison, each of the problems is solved with the following four methods:

\textbf{Method A:} GGN with \textit{central finite difference} approximation of the Jacobian of $R$ in mode with small perturbations $\epsilon\approx1.49\times10^{-8}$ \cite[Eq. 1.2]{Berahas.2022}.  \\
   \textbf{Method B:} Gradient Descent.  
    The gradient $\nabla C$ of the objective Eq.~\eqref{eq_objective}  is approximated using finite differences and determines the step in the descent direction.\\
     \textbf{Method C:} Deterministic MPPI, as presented in Sec.~\ref{sec:mppi}.\\
    \textbf{Method D:} Gauss-Newton accelerated MPPI (cf. Algo.~\ref{alg:GNaccMPPI}) with random sampling of Eq.~\eqref{eq_Gaussian_smoothing_jacobian}.\\
    \textbf{Method E:} Gauss-Newton accelerated MPPI (cf. Algo.~\ref{alg:GNaccMPPI}) with deterministic \textit{Dirac-mixture} approximation of Eq.~\eqref{eq_Gaussian_smoothing_jacobian}. This technique is based on the sigma-point approach utilized in the Unscented Kalman Filter (UKF) \citep{Wan2000}.  

 For Methods {C} and {D}, $M=2000$ samples with $\lceil M/M_\mathrm{GPU} \rceil=1$ are used, and antithetic sampling is employed to improve performance, cf. Sec.~4.2 in \cite{Glasserman.2003}.  For reproducibility, the full implementation is provided in the repository: \texttt{tba -- after publication}. 

\subsection{Results}
The results obtained by applying the different methods to the control problems are presented in Table \ref{tab:MPPI_comparison_results}. Each entry in the table is encoded as
 \begin{equation*} 
    \left[\begin{array}{c}
        \{1,\ldots,1000\}  \\
        \{\checkmark,-,\times \} 
    \end{array}\right]\equiv
      \left[\begin{array}{c}
         \mathrm{number\; of\; iterations}  \\
         \{\mathrm{opt.\; sol., subopt.\; sol., no\; sol.} \} 
    \end{array}\right].
\end{equation*}
The optimization is stopped if the step length and the (smoothed) gradient of the objective fall below a small threshold.
 The iteration count is a suitable indicator of computational effort, since each method requires two sequential evaluations of the inner function.

\begin{table}[t]
\medskip
 	\begin{center}
 		\caption{Results of the numerical experiments. }
 		\label{tab:MPPI_comparison_results}
 		\begin{tabular}{lccccc}
         \hline \hline
 		Method&   A &B  &  C&  D& E \\
 		\hline
 		Pr. I   &\cellcolor[HTML]{D7D8EC} $\begin{array}{c}
 		      \textbf{12}\\
 		     \textbf{\checkmark}
 		\end{array}$ &$\begin{array}{c}
 		      1000\\
 		     \times
 		\end{array}$ &$\begin{array}{c}
 		      17\\
 		     \checkmark
 		\end{array}$ & $\begin{array}{c}
 		      14\\
 		     \checkmark
 		\end{array}$
        & \cellcolor[HTML]{D7D8EC} $\begin{array}{c}
 		     \textbf{12}\\
 		     \checkmark
 		\end{array}$\\
\hline        
 		Pr. II  & $\begin{array}{c}
 		      3\\
 		     -
 		\end{array}$ &$\begin{array}{c}
 		      4\\
 		     -
 		\end{array}$ &$\begin{array}{c}
 		      7\\
 		     \checkmark
 		\end{array}$ & $\begin{array}{c}
 		      4\\
 		     \textbf{\checkmark}
 		\end{array}$
        & \cellcolor[HTML]{D7D8EC} $\begin{array}{c}
 		      \textbf{2}\\
 		     \checkmark
 		\end{array}$\\
\hline        
 		Pr. III  & \footnotesize $\begin{array}{c}  
 		      \mathrm{not}\\ \mathrm{admissible}
 		\end{array}$ &\footnotesize $\begin{array}{c}  
 		      \mathrm{not}\\ \mathrm{admissible}
 		\end{array}$  &\cellcolor[HTML]{D7D8EC} $\begin{array}{c}
 		      \textbf{2}\\
 		    \textbf{ \checkmark}
 		\end{array}$ &\cellcolor[HTML]{D7D8EC} $\begin{array}{c}
 		     \textbf{2}\\
 		    \textbf{ \checkmark}
 		\end{array}$ &\cellcolor[HTML]{D7D8EC} $\begin{array}{c}
 		     \textbf{2}\\
 		    \textbf{ \checkmark}
 		\end{array}$\\
\hline        
        Pr. IV & \cellcolor[HTML]{D7D8EC} $\begin{array}{c}
 		      \textbf{2}\\
 		    \textbf{\checkmark} 
 		\end{array}$ &$\begin{array}{c}
 		       42\\
 		    \checkmark 
 		\end{array}$ &$\begin{array}{c}
 		      45\\
 		     \checkmark
 		\end{array}$ & $\begin{array}{c}
 		      7\\
 		    \checkmark 
 		\end{array}$ &\cellcolor[HTML]{D7D8EC} $\begin{array}{c}
 		     \textbf{2}\\
 		    \textbf{ \checkmark}
 		\end{array}$\\
        \hline        
        Pr. V.i  & \cellcolor[HTML]{D7D8EC}  $\begin{array}{c}
 		      \textbf{3}\\
 		    \textbf{\checkmark}
 		\end{array}$ &$\begin{array}{c}
 		      681 \\
 		    \checkmark 
 		\end{array}$ &$\begin{array}{c}
 		      48\\
 		    \checkmark 
             
 		\end{array}$ & $\begin{array}{c}
 		       5\\
 		    \checkmark 
 		\end{array}$&\cellcolor[HTML]{D7D8EC}  $\begin{array}{c}
 		      \textbf{3}\\
 		    \textbf{\checkmark}
 		\end{array}$ \\
        \hline        
        Pr. V.ii  &  \footnotesize $\begin{array}{c}  
 		      \mathrm{not}\\ \mathrm{admissible}
 		\end{array}$  &\footnotesize $\begin{array}{c}  
 		      \mathrm{not}\\ \mathrm{admissible}
 		\end{array}$  &$\begin{array}{c}
 		      50\\
 		      \checkmark
 		\end{array}$ & $\begin{array}{c}
 		       6\\
 		      \textbf{\checkmark }
 		\end{array}$ & \cellcolor[HTML]{D7D8EC}$\begin{array}{c}
 		       \textbf{3}\\
 		      \textbf{\checkmark }
 		\end{array}$ \\
 		    \hline \hline
 		\end{tabular}
 	\end{center}
 \end{table}
 
The results show that the GGN method with finite differences (Method A) achieves fast convergence in well-posed problem settings such as the Rosenbrock function and linear or nonlinear MPC with smooth objective functions. However, it converges only to a suboptimal solution for the Rastrigin function and is not admissible for optimizing nonsmooth problems with vanishing gradients. The gradient descent method with finite differences (Method B) generally requires a large number of iterations to converge (see, however, Problem II). It fails to solve the Rosenbrock function within 1000 iterations.

In contrast, the deterministic MPPI method (Method C) and the Gauss-Newton accelerated MPPI methods (Variant D and E) converge across all tested problem classes. In particular, Method E consistently achieves faster convergence (up to $\times$21 fewer iterations), especially in high-dimensional settings (cf. Problems IV, V.i, and V.ii) compared to deterministic MPPI (Method C). Note that the deterministic MPPI method requires 45 iterations for the linear optimal control example (Problem IV), even though this problem is a quadratic program that could be solved in a single Newton step. Remember the limitation of the presented Gauss-Newton accelerated MPPI approach given by the restriction to the convex-over-unknown problem structure \eqref{eq:MPPI_convex_over_non}.  However, as previously discussed, this structure is widespread in optimal control applications. 

\section{Conclusion}\label{sec:con}
 This paper introduces the \textit{Gauss-Newton accelerated MPPI method}, which combines the complementary strengths of sampling-based and second-order optimization techniques. The numerical results demonstrate that the flexibility of MPPI for solving black-box problems is retained, while the rapid convergence properties of the Generalized Gauss–Newton (GGN) method can be effectively exploited. The proposed approach is well-suited for high-dimensional, unconstrained optimal control problems with a \emph{convex-over-unknown} objective structure. This setting commonly arises when system dynamics are represented by black-box simulation engines. %Future work could extend the presented approach to constrained optimization problems.

\begin{ack}
	
The authors express their sincere gratitude to Prof. Dr. Oliver Dürr for our discussions on Monte Carlo methods.
\end{ack}

\bibliography{references}

%\bibliography{ifacconf}             % bib file to produce the bibliography
                                                     % with bibtex (preferred)

%\appendix
%\section{A summary of Latin grammar}    % Each appendix must have a short title.
%\section{Some Latin vocabulary}              % Sections and subsections are supported  
                                                                         % in the appendices.
\end{document}